\documentclass[a4paper,longnamesfirst,useAMS,usenatbib,usegraphicx,usedcolumn]{mn2e}

\usepackage{amsmath}
\usepackage{times}

\usepackage{url}
\usepackage{xr-hyper}
\usepackage{hyperref}

\shortcites{browne93,biggs02,tammann01,parodi00,press_etal,cohen00,biggs01,kochanek89,patnaik93,biggs99}

\newcommand{\nt}{}
\newcommand{\papi}{Paper~I}
\newcommand{\papii}{Paper~II}

\newcommand{\mat}[1]{{\mathbfss{#1}}}

\DeclareMathOperator{\trace}{Tr\,}

\newcommand{\vc}[1]{\bmath{#1}}
\newcommand{\ft}[1]{\tilde{#1}}
\newcommand{\diff}{{\rmn{d}}}

\newcommand{\dii}[1]{\! \diff^2\!#1\,}
\newcommand{\tpi}{2\upi\,{\rmn{i}}}
\newcommand{\sprod}{\cdot}
\newcommand{\expa}[1]{{\rmn{e}}^{#1}}

\newcommand{\transp}[1]{{#1}^{\dagger}}
\newcommand{\ID}{I\sub D}
\newcommand{\vcID}{\vc{I}\sub D}

\newcommand{\kmsmpc}{\ifmmode\text{km}\,\text{s}^{-1}\,\text{Mpc}^{-1}\else$\text{km}\,\text{s}^{-1}\,\text{Mpc}^{-1}$\fi}

\newcommand{\mymathstrut}{jkl'm}

\newcommand{\subm}[1]{_{#1\vphantom{\mymathstrut}}}

\newcommand{\sub}[1]{_{\mathrm{#1}}}

\newcommand{\conj}[1]{{#1}^{\star}}

\newcommand{\mas}{{\ifmmode\text{mas}\else mas\fi}}

\newcommand{\BOii}{B0218+357}

\newcommand{\rtext}[1]{\quad\text{#1}}

\newcommand{\lc}{{\scshape LensClean}}
\newcommand{\lcc}{{\scshape (Lens)Clean}}

\newcommand{\lensmem}{{\scshape LensMEM}}

\newcommand{\lentil}{{\scshape LenTil}}

\newcommand{\kne}{{\scshape KNE}}

\newcommand{\knelc}{\kne-\lc}

\newcommand{\clean}{{\scshape Clean}}

\newcommand{\nnls}{{\scshape NNLS}}

\newcommand{\difmap}{{\scshape DIFMAP}}

\newcommand{\aips}{{\scshape AIPS}}

\let\epsilon=\varepsilon

\title[LensClean revisited]{\lc\ revisited}
\author[O.~Wucknitz]
       {O.~Wucknitz,$^{1,2,3}$\thanks{E-mail: {\tt
  olaf@astro.physik.uni-potsdam.de}} 
  \\
$^1$ University of Manchester, Jodrell Bank Observatory, Macclesfield, Cheshire SK11 9DL, UK
         \\
$^2$ Hamburger Sternwarte, Universit{\"a}t Hamburg, Gojenbergsweg 112, 21029 Hamburg, Germany
\\
$^3$ Universit\"{a}t Potsdam,   Institut f\"{u}r Physik, Am Neuen Palais 10,
  14469 Potsdam, Germany
}

\begin{document}

\maketitle

\begin{abstract}
We discuss the \lc\ algorithm which for a given gravitational lens model fits
a source 
brightness distribution to interferometric radio data in a similar way as
standard \clean\ does in the unlensed case. The lens model
parameters can then be varied in order to minimize the residuals and determine
the best model for the lens mass distribution.
Our variant of this method is improved in order to be useful and stable
even for high dynamic range systems with nearly degenerated lens model
parameters. Our test case \BOii\ is dominated by two bright images but the
information needed to constrain the unknown parameters is
provided only by the 
relatively smooth and weak Einstein ring. The new variant of \lc\ is able to
fit lens models even in this difficult case.
In order to allow the use of general mass models with \lc, we develop the new
method \lentil\ which inverts the lens equation much more reliably
than any other method. This high reliability is essential for the use as part
of \lc.
Finally a new method is developed to produce source plane maps of the unlensed
source from the best \lc\ brightness models. This method is based on the new
concept of `dirty beams' in the source plane.

The application to the lens \BOii\ leads to the first useful constraints for
the lens position and thus to a result for the Hubble
constant. These results are presented in an accompanying \papii, together with
a discussion of classical lens modelling for this system.
\end{abstract}


\section{Introduction}

The subject of gravitational lensing has matured from a curiosity and test of
the theory of general relativity to an invaluable tool for a wide area of
astrophysical applications, ranging from cosmology down to the study of compact
objects within our own galaxy.
One of the most important cosmological applications is the determination of the
Hubble constant from time-delays between multiple images of lensed
extragalactic sources as proposed by \citet{refsdal64b}. The idea is simple:
By measuring image positions and making model assumptions for the mass
distribution, the general geometry of a given lens system can be determined.
With the determination of only one length in this geometry, all other lengths,
especially the distances to the lens and source, can immediately be deduced,
thus allowing the determination of $H_0$ from the measured redshifts.
The essential length can be provided by measurements of the light travel time
difference (`time-delay') between two images of the same source.
The method does not rely on the understanding of complicated astrophysical
processes but only on the validity of the static weak-field limit of
general relativity, which is confirmed by numerous tests, and on the
Friedmann--Robertson-Walker cosmological model.

The only critical topic is the mass distribution of the lens in
question. Using simple models like singular isothermal ellipsoidal mass
distributions, the constraints provided by the image geometry are sufficient
to determine the model parameters for many lens systems with high accuracy.
Unfortunately results obtained in this way are very model-dependent and may
thus be biased by astrophysical prejudice.
The geometry of multiply imaged point sources can naturally not provide a
sufficient number of 
constraints to determine the parameters of more general model families.
In order to overcome the difficulty, more information has to be included
in the modelling. This can be achieved by using multiply imaged
\emph{extended} sources in which each sub-component of the source provides its
own set of constraints.

Extragalactic sources do generally show more structure at
radio wavelengths than in the optical, which is an important motivation to
concentrate on radio lenses. Additionally the use of radio
interferometers of different sizes at different wavelengths allows the study on
scales from below milli-arcseconds to above arcminutes. Even with the most
modern telescopes this is not possible at optical wavelengths.

In the case of well-resolved and separated multiple subcomponents, the
standard approach of first measuring the positions, flux densities and shapes
of the subcomponents and using these parameters for the modelling works very
well. In the general case, however, the structure of radio sources cannot be
parametrized in a simple way.
Several approaches for models using extended emission without parametrizing
the source
have been proposed in the literature which all follow the same general
concept. For a given lens model they construct a map of the source
which minimizes the deviations from observations when mapped back to
the lens plane. The minimal residuals themselves are then used in an
outer loop to find the best lens model.

The first and most simple method we tried is the so called ring cycle
\citep{kochanek89}. It is based on pixellated maps of the true lensed surface
brightness of the system. For each pixel in the source plane, a mean
of corresponding surface brightnesses in the lens plane is
calculated. The scatter in these pixels can be combined to get a
meaningful measure for the deviation of the observations from the best
model map. The weakness of this algorithm is its dependence on the
true surface brightness. Optical and radio data always provide
maps which are a convolution of the true brightness distribution with
the point spread function (optical) or the dirty beam (radio). One
might think of deconvolving these maps to get a more accurate estimate
of the true image, but this process always introduces artifacts and
biases which will show as difficult to interpret errors in the final
results. This is a typical inverse problem and it is much better to
apply the well understood effects of the observational process on the
model data to compare directly with the observations than to do it the other
way round and try to
correct the observations to compare with the model.

For the problem of lens modelling with radio data, this leads directly
to the \lc\ algorithm which was introduced by \citet{kochanek92}
and \citet{ellithorpe96}. We do not use \lensmem\ \citep{wallington96}, which
is a formulation of 
the classical maximum entropy method for a lensed situation, because the
regularisation with an entropy term changes 
the residuals in a way which is difficult to interpret statistically. Since we
rely on the residuals to determine the best lens models, we prefer to avoid
such effects.

In this paper we describe a number of improvements of the original \lc\
method. The motivation to start the development of our own implementations was
the study of the radio lens system \BOii\ \citep{patnaik93} which has a
measured time-delay \citep{biggs99,cohen00} and seems to be especially
well-suited for Refsdal's method of determining $H_0$.
The most serious problem in this lens system is the small size (the separation
of the two images of ca.\ 330\,mas is the smallest of all known lenses) which
makes direct optical measurements of the lens position very difficult. Because
the asymmetry of the geometry, which causes the time-delay, does directly
depend 
on the lens position, no results are possible without a good estimate for this
parameter.
The two images of \BOii\ show substructure on milli-arcsecond scales
\citep{patnaik95,kemball01,biggs02} and an Einstein ring
of the same size as the image-separation \citep[e.g.][]{biggs01}.
It is the rich structure of this ring which potentially provides good
constraints for the lens position, while the substructure of the images is
more valuable for the radial mass profile.

\lc\ in its original form has serious
shortcomings which prevent its use in a system like \BOii, where the dynamic
range between the compact images and the ring is very high and the important
model constraints are provided by the weaker components.
The discussion of our significantly improved version of  \lc\ comprises the
main part of this article.
This new version will in \papii\ be used to determine the galaxy
position with an accuracy that is sufficient to achieve a result for
the 
Hubble constant which is competitive with other methods but avoids their
possible systematical errors.

As a basis for the reconstruction of the true source brightness distribution we
will 
generalize the concept of the `dirty beam' from the usual image plane
formulation to the  source plane. Certain approximations will lead to
a relatively simple but well founded recipe to restore the unlensed source
plane from \lc\ brightness models.

The details of \lc\ and our modifications and improvements of it are described
in this article (\papi) while the classical lens modelling and the
\emph{results} of \lc\ for \BOii\ are discussed in \papii\ \citep{paper2},
appearing in the same issue of this journal.
Both papers are condensed versions of major parts of \citet{phd}. For many
more details, especially about the development of our version of \lc, the
reader is referred to that work.

\section{The test case: Data and models}

Most of the development and tests have been performed with a 15\,GHz data
set of \BOii\ taken with 26 antennas of the VLA in A configuration in 1992 as
part of program AB~631 (P.I. Ian Browne).
The long-track observations were done in full polarization at 14.965\,GHz with
a bandwidth of 50\,MHz and a total on-source time of slightly less than 6
hours. The initial 10~sec integrations were further binned to 1~min in our
calculations to reduce the amount of data and the computation times.

The data have been calibrated in a standard way including mapping and
self-calibration with \aips\ and \difmap.
The beam size is $129\times146\,\mas^2$ (p.a. $-73\,\degr$) with natural
and $86\times 88 \,\mas^2$ (p.a. $-60\,\degr$) with uniform weighting.
The thermal rms noise per beam is 55\,$\umu$Jy for natural and
140\,$\umu$Jy for uniform weighting.

For the lens models we use the standard approach of elliptical power-law
potentials described in \papii. The radial part of these potentials is
$\psi\propto r^\beta$ and most of the work so far has been done for the
isothermal case $\beta=1$, using `singular isothermal elliptical potentials'
(SIEP models). Classical modelling using VLBI constraints shows
that the deviations from isothermality are small ($\beta\approx1.04\pm0.02$,
cf.\ \papii) and do not affect most of the work on the \lc\ algorithm.
To be able to use non-isothermal lens models with \lc\ in the future, we had
to invent the new method \lentil\ to solve the lens equation which will be
discussed briefly in this paper.

\section{Radio interferometry}

Before coming to \lc\ itself, we first have to discuss
how radio interferometers
measure components of the Fourier transform of the true
brightness distribution on the sky and how these can be used to infer the
source brightness distribution by using the \clean\ algorithm.
More details can be found in standard text books on the subject
\citep[e.g.][]{thompson86,perley86,perley89,taylor99}. 

All lens systems are very small
compared to the size of the celestial sphere, making it possible to
approximate the sky by a tangential plane and use two
dimensional Fourier transforms.
We call the angular sky coordinates $\vc z=(l,m)$; the true brightness
distribution is $I(\vc z)$. Coordinates in Fourier space are $\vc
u=(u,v)$. Each pair of telescopes measures one component of the
Fourier transform $\ft{I}$ for each integration bin, called a
visibility $\ft{I}_j$ for the baseline $\vc u_j$.
\begin{equation}
\ft I (\vc u_j) = \int\dii z I(\vc z)\, \expa{\tpi\,\vc u_j\sprod\vc
  z} + \ft{R}_j
\label{eq:ft radio}
\end{equation}
Here the thermal noise of the receivers, atmosphere etc.\ is denoted
by $\ft{R}_j$. 
The inverse Fourier transform of only the \emph{measured} spatial frequency
components is called the `dirty map'
\begin{equation}
\ID (\vc z) = \frac{1}{W}\sum_j w_j\, \ft{I}_j\, \expa{-\tpi\,\vc u_j\sprod\vc
  z} \rtext{,} 
\nt
\end{equation}
where $w_j$ is a weighting function. See \citet{briggs99} for a discussion of
 different weighting strategies. The scaling by $W=\sum_j w_j$ is
done to normalize physical units concordant with $\ft I$ (usually Jy).
For so called natural
weighting, the weights are the formal statistical weights of the
visibilities, $w_j=\sigma_j^{-2}$. For uniform weighting, the weights are
chosen to achieve a 
constant density of weights in the areas of the $uv$ plane where
measurements exist.

Some care is necessary here to count the visibilities correctly. The variance
$\sigma_j^2$ of the visibilities is defined to be the variance of either real
or imaginary part. The effective number of measurements $\nu$ is \emph{twice}
the number of visibilities because real and imaginary parts are counted
separately. A somewhat more elegant approach is to include for each measured
visibility $\ft{I}(\vc u_j)$ also the reflected value for $-\vc u_j$, which has
the value $\conj{\ft{I}}(\vc u_j)$, to obtain a measurement set with its
natural symmetries. The total number of (true and reflected) visibilities is
then $\nu$. To keep the error statistics correct, the weights $w_j$ have to be
shared between true and reflected visibilities.% which halves there values.
With this symmetrical completion, the dirty beam (see below) and dirty map
always become real. This approach makes the further analytical calculations
much simpler 
because it avoids the need to select real parts of complex quantities at
several occasions.

Even without taking into account measurement errors, the dirty map
does not represent the true brightness distribution but a convolution
of the latter with the `dirty beam' $B$.
\begin{equation}
B(\vc z) = \frac{1}{W}\sum_j w_j \, \expa{-\tpi\,\vc u_j\sprod\vc z}
\nt
\end{equation}
The problem of deconvolving the map is not solvable uniquely because
of incomplete coverage of $uv$ space. We will not discuss optimal strategies
to solve this
problem here because we are only interested in \emph{one} of the optimal
solutions (which all have equal residuals) and in the residuals themselves in
particular. Whether this solution is the 
`best' one in the sense of the most probable brightness distribution of a real
physical source is of secondary importance in our context.

Direct inversion of the convolution equation, which is a system of linear
equations, is possible but 
numerically expensive. The standard method to get at least an
approximate map is the \clean\ algorithm \citep{hoegbom74}. It works by
successively subtracting fractions of the highest peaks in the map from the
visibility data and converges to one of the best possible solutions in
the infinite limit. Convergence is fast in the beginning but becomes very slow
in the later stages when smooth surface brightness dominates
the residuals. Details of the \clean\ algorithm will be discussed later as a
special case of \lc. The mathematical foundation for the
heuristic \clean\ method was laid by \citet{schwarz78}.

Practical computation of the Fourier transforms is usually done by
a fast Fourier transform (FFT) of gridded data. To minimize the effect
of aliasing (folding of emission outside of the map area into the map as an
effect of the regular grid), the data are convolved in $uv$ space with a
smoothing kernel, the 
effect of which is corrected for after the FFT, by dividing by the inverse FT
of the convolution function. In this way the response to emission outside of
the map is reduced dramatically. Details of this standard approach can be
found in \citet{briggs99}.

\section{The LensClean algorithm}  % \lc

\lc\ was first proposed by \citet{kochanek92} and later improved
by \citet{ellithorpe96}. We present a simple while more general analysis
of this method.
In standard \clean, components to be removed can be selected freely,
with the only constraint of being located in windows outside of which
no emission is expected. In \lc, for every position in the
source plane, emission has to be subtracted at the positions of \emph{all}
corresponding image positions simultaneously with the
magnifications\footnote{We use the term `magnification' throughout this
paper. Lensing conserves surface brightness so that this magnification shows
as `amplification' for pointlike components.}
given by the (for now fixed) lens model. In standard \clean, the next
peak to subtract always is the pixel with highest flux density in the
map. This leads to the highest possible decrease of the residuals in
this particular iteration.
To generalize for a lensing scenario, we still insist on steepest
decline of the residuals in each step because of the success of this
approach in non-lensed \clean.

\subsection{Standard variant (\kne)}

Let us calculate the residuals after subtracting the images $k=1\dots n$
corresponding 
to a certain component in the source plane with flux density
$S$. Positions and magnifications of the images are
$\vc z_k$ and $\mu_k$. These as well as the number of images $n$ to
include depend on the lens model and source position. The weighted sum of
squares of the residual visibilities is 
\begin{align}
R^2 &=  \sum_j w_j \left| \ft{I}_j - S \sum_k \mu_k\expa{\tpi\,\vc
    u_j\sprod \vc z_k} \right|^2 \nt \\
&= \sum_j w_j | \ft{I}_j |^2 -W\Delta \rtext{,}
\nt
\\
\Delta &= 
2 \, S \, \sum_k \mu_k I_k-
S^2\, \sum_{kk'} \mu\subm{k} \mu\subm{k'} B_{kk'} 
\rtext{.} %\nt
\label{eq:R^2 2}
\end{align}
where we used the following definitions for the components of dirty beam and
map: 
\begin{align}
B_{kk'} &= B (\vc z\subm{k}-\vc z\subm{k'}) \nt \\
I_k &= \ID (\vc z_k) \nt
\end{align}
\citet{kochanek92} used a different approach and calculated the residuals in
 image space which is equivalent to applying the weights quadratically to the
 $uv$ space residuals.
Since the measurements are done in $uv$ space,
it is preferable to calculate the residuals directly on these data.
This was also done by \citet{ellithorpe96} with the restriction to
naturally weighted data. Our derivation of \lc\ is valid for all
weighting schemes in a statistically rigorous way with no approximations
necessary. It is equivalent
to \citet{kochanek92} only for uniform weighting.

To get the minimal residuals for the fixed source position, the
derivative of $\Delta$ with respect to $S$ has to vanish, leading to a source
flux density of
\begin{equation}
S = \frac{\sum\limits_k \mu_k I_k}{\sum\limits_{kk'}
  \mu\subm{k} \mu\subm{k'} B_{kk'}} \rtext{.}
\label{eq:S KNE}
\end{equation}
Now we have to find the source position maximizing the residual difference
caused by the subtraction of the flux density just calculated. We therefore
have to find the maximum of 
\begin{align}
\Delta &= \frac{\Bigl(\sum\limits_k
  \mu_k I_k\Bigr)^2}{\sum\limits_{kk'} \mu\subm{k} \mu\subm{k'} B_{kk'}} \rtext{.}
\label{eq:delta}
\end{align}
To stabilize the
algorithm and to accelerate convergence in later stages, we do not
subtract the total flux calculated but a fraction $\gamma S$ using a loop gain
$\gamma$ of the order $0.1$. The value of $\gamma$ does not influence the
selection of the source position.
We refer to this variant as `KNE' \lc\ (standing for the authors of
\citealp{kochanek92} and \citealp{ellithorpe96}). 

The special case of an unlensed source is equivalent to the
standard \clean\ algorithm. In this situation, the optimal position and flux
of the next component are given by the peak (in the sense of largest absolute
value) of the residual map $I$.

\subsection{New unbiased variant}

When testing the \knelc\ method as described above with simulated data, we
noticed a serious shortcoming which introduced systematic errors into
the results. The residuals were not a continuous function of the lens
model but showed jumps especially at places in parameter space where regions
of higher multiplicities appear in the system.

This standard choice of \clean\ components is appropriate for well
separated point sources but not for smooth surface brightness sources.
Consider a well resolved source with a constant true surface brightness and
therefore constant observed surface brightness $I(\vc x)$ in the image plane. 
There are two reasons to modify the selection of \lc\ components
in this case. For efficient \clean{}ing, components have to be
subtracted evenly distributed over the area of constant surface
brightness. 
Furthermore we must avoid any bias for certain lens models
which are equivalent with others in respect of capability to explain
the observations. The decrease in residuals $\Delta$ from
Eq.~\eqref{eq:delta}
is, on the other hand, clearly not independent of lens model and source
position as required. Note that
\emph{all} lens models are equally compatible with the constant surface
brightness scenario.

\begin{figure}
\includegraphics[width=8cm]{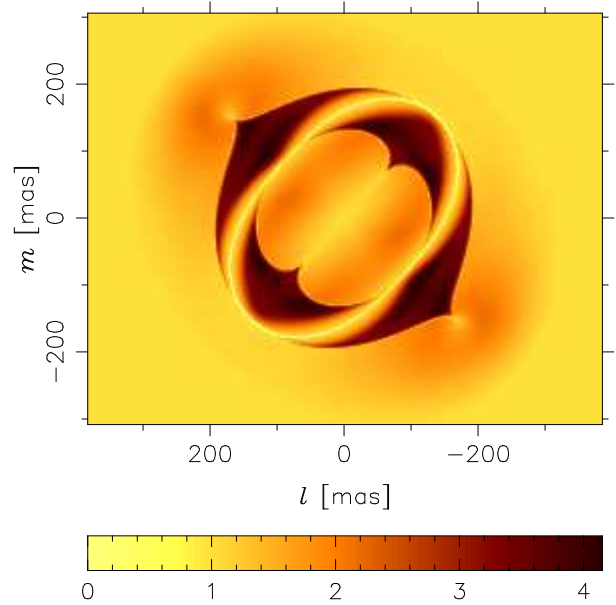}
\caption[Bias factor in the KNE \lc\ variant]{Bias factor in the KNE \lc\
  variant. This factor is proportional to 
  the residual reduction from Eq.~\eqref{eq:delta} in one \lc\ iteration for a
  constant   surface 
  brightness source. The lens model assumed for this plot is very similar to
  the best model for   \BOii.} 
\label{fig:bias}
\end{figure}

We tried different methods of improving \lc\ in this respect, the
most successful one we want to present here.
The simple idea is to apply an adaptive loop gain
depending on source position and lens model.
This is applied as a factor to $\Delta$ instead of $S$. To obtain a unique
recipe and keep things simple,
we demand that this factor does not depend on the
observed brightness distribution.
We can then assume an arbitrary brightness distribution to calculate the
optimal correction factor and use the constant surface brightness scenario for
this. 
The correction factor is then simply the inverse bias factor we would get from
Eq.~\eqref{eq:delta} in this case (see also
Fig.~\ref{fig:bias}):
\begin{align}
\Delta' &= 
g\,\frac{\sum\limits_{kk'} \mu_k \mu_{k'} B_{kk'}}
{\Bigl(\sum\limits_k \mu_k\Bigr)^2} \, \Delta \nt \\
&=
g\left(\frac{\sum\limits_k
  \mu_k I_k}{\sum\limits_k \mu_k}\right)^2  \rtext{.} \nt
\end{align}
We included a factor $g$ here which plays the role of a loop gain scaling
factor  and in the case
of no lens is 
the same as $\gamma\,(2-\gamma)$ and therefore monotonically related to
the conventional loop gain $\gamma$ for $0\le\gamma\le1$.
We now select the source position for the next iteration by searching
the maximum of $\Delta'$. The flux density which should be subtracted at this
position 
to achieve exactly this decrease in $R^2$ can easily be calculated by going
back to Eq.~\eqref{eq:R^2 2} and solve for the new $S'$ with the
known $\Delta'$. 
\begin{equation}
S' =
 \left(
  1-\sqrt{1-g\,\frac{\sum\limits_{kk'}\mu_k\mu_{k'}B_{kk'}}{\Bigl(\sum\limits_k
  \mu_k\Bigr)^2}}\right) \, S
\label{eq:S unbiased}
\end{equation}
This expression can be interpreted more easily in the limiting case
of a small gain $g$:
\begin{equation}
S' \approx \frac{g}{2}\, \frac{%\displaystyle
\sum\limits_k \mu_k I_k
  }{
\left(\sum\limits_k \mu_k\right)^2}
\nt
\end{equation}
This is proportional to a mean of the source plane fluxes $I_k/\mu_k$,
estimated from the 
individual images, weighted with $\mu_k^2$ and scaled with $g/2$.
The remaining scaling factor is responsible for compensating the bias.

Since in \BOii\ we have not only a relatively well resolved ring but
also the two strong compact components, we have to assure that the
modified method also works in this case. For images with equal
magnification, the adaptive loop gain corrects for the bias caused by
the number of images, which is sensible.
Numerical tests show that the adaptive gain
works very well for compact and extended emission and minimizes the bias
effects.
Due to the different definition, higher values can be
used for $g$ than for $\gamma$ without deteriorating the results. For
equal magnifications of $n$ well separated images, both methods are
equivalent if 
\begin{equation}
g = n\,\gamma\,(2-\gamma)
\nt
\end{equation}
is used.

Please note that the selection rule of \lc\ components does not influence
the results in the idealized case where
the \clean\ iterations are performed until convergence is reached. For
practical work, however, convergence of \lcc\ is so slow at later stages that
the unbiased \lc\ variant actually \emph{does} improve the results
considerably even if 
a few 1000 iterations are done. The real
differences caused by different lens models are so small that any bias effects
have to be avoided by all means. We therefore used our unbiased \lc\ variant
for all the calculations presented here.

The more uniform \clean{}ing with the unbiased algorithm does also accelerate
convergence for good lens models by an order of magnitude while it does not
influence the residuals in 
other cases very much. It therefore improves the ability of the residuals
to discriminate between good and bad
lens models after a moderate number of iterations.

\begin{figure*}
\includegraphics[width=0.9\textwidth]{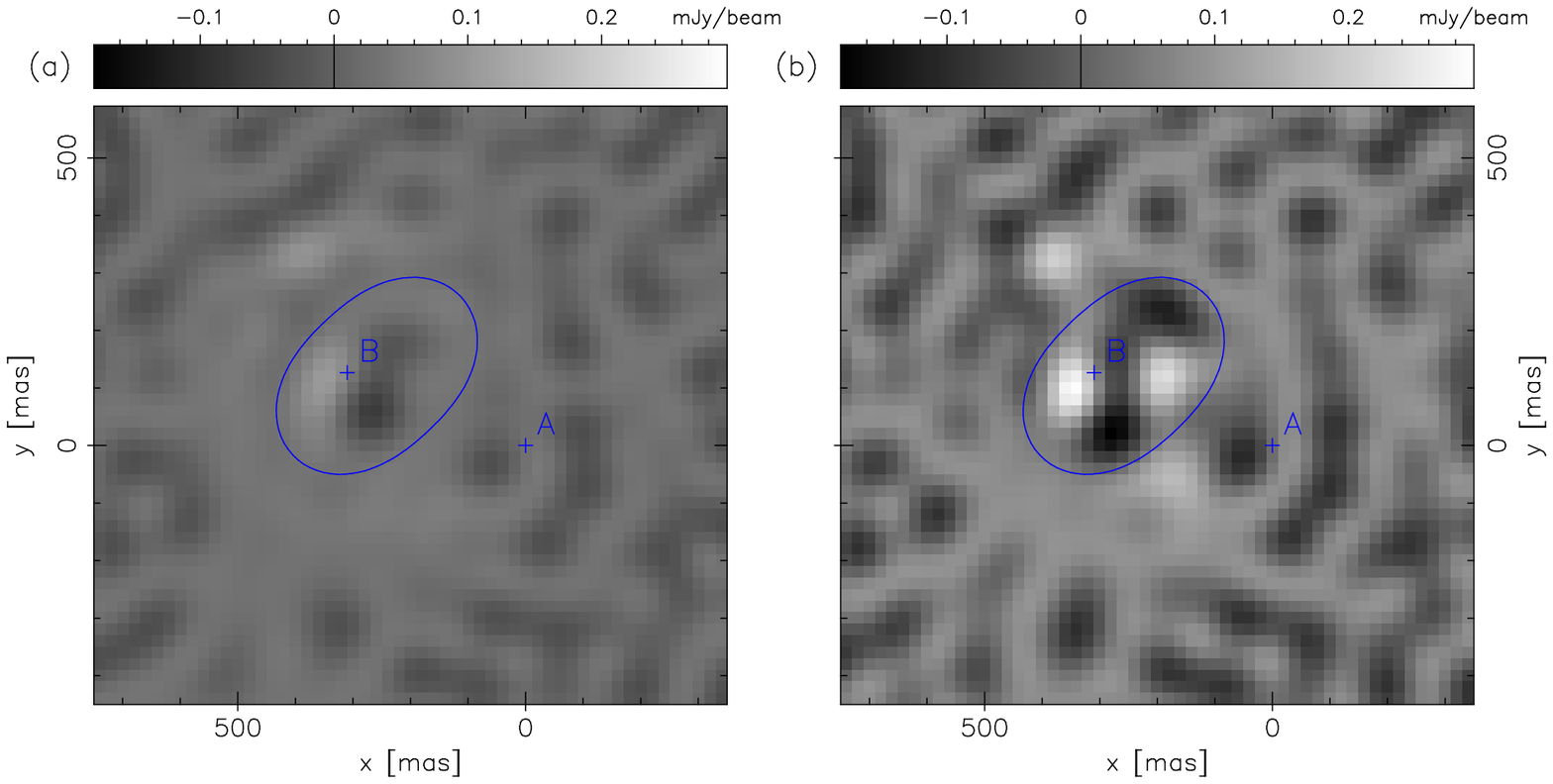}\\[2ex]
\includegraphics[width=0.9\textwidth]{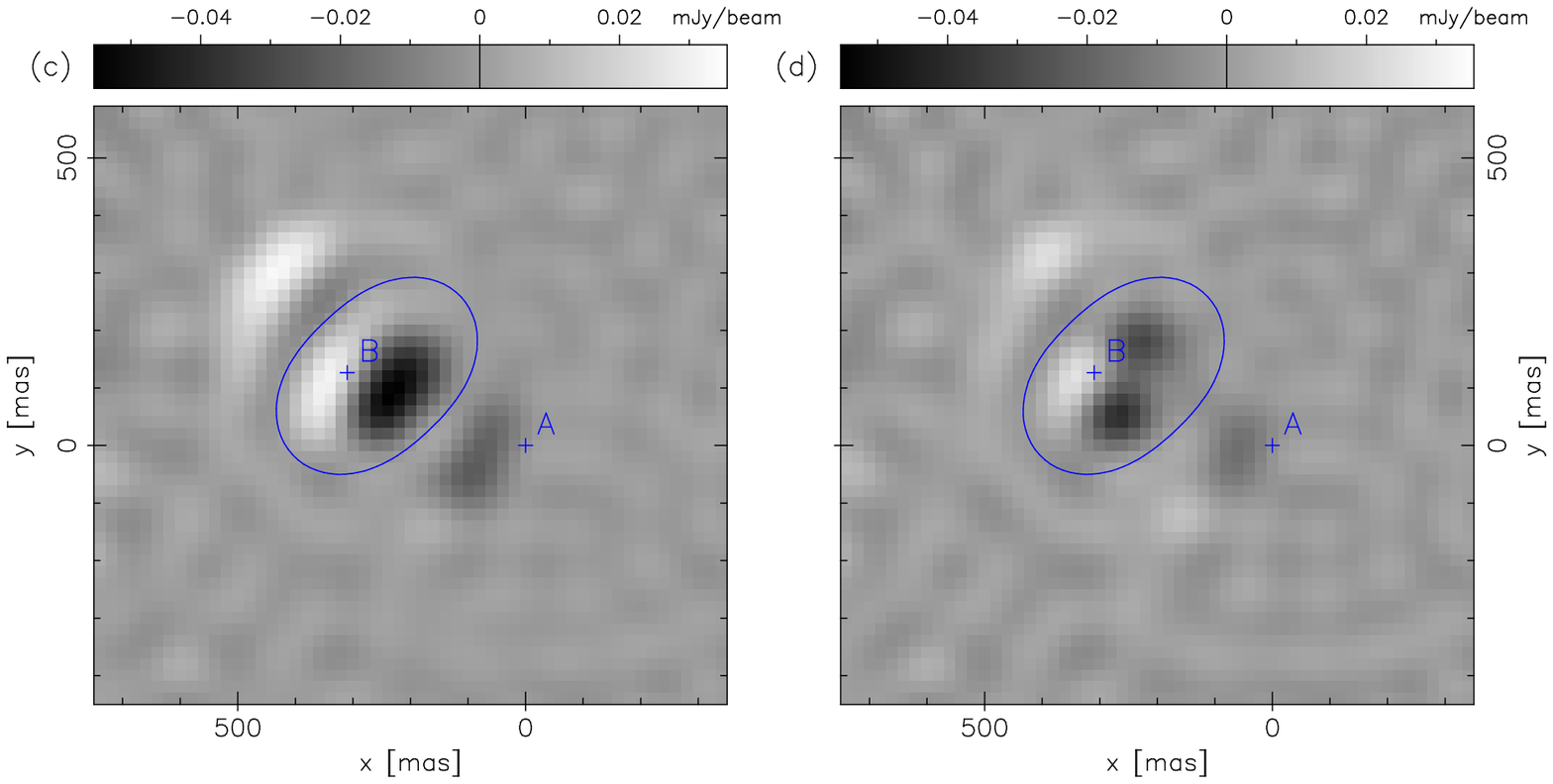}
\caption[Simulated image plane residuals]{Uniformly weighted image plane
  residuals for a 
  simulated data set using the known lens 
  model parameters. The top panels (a+b) show the results after 2000, the
  bottom panels (c+d) after 100\,000 iterations (with a very different grey
  scale; the total range shown is a factor of $\sim5$ smaller). The unbiased variant
  is on the  
  left (a+c), the \kne\ variant on the right (b+d).
  For 2000 iterations the unbiased version of \lc\ is clearly superior while
  both are more or less equivalent in the limit of very many iterations. The
  expected noise 
  of the dirty maps is 0.14\,mJy per beam.
  The critical curve and the two bright images are marked.
}
\label{fig:resid1}
\end{figure*}

In Fig.~\ref{fig:resid1} we compare image plane residuals for both variants of
\lc. For these maps we used a simulated data set which was made using the
best fitting lens and source model for \BOii. After building the $uv$ data set
with the 
same $uv$ coverage as in the real data set, we added noise and performed
self-calibration with the known emission model. This last step is necessary to
be able to compare the results with \lc\ runs for the real data set which will
be presented in \papii.
We see that for a moderate number of 2000~iterations, which is quite typical
for real model fitting runs, the residuals are significantly higher for the
\kne\ variant than for the new unbiased version. The alignment of the
residuals with the critical curve of the lens could in such a case be
misinterpreted as a bad fit of the lens model.
Only if many more iterations are performed do the residuals become similar.

Note that the image space residuals become much smaller than the expected rms
noise in the 
dirty map. This is well known from unlensed \clean\ where the residuals, by
including the noise in the emission model, can be reduced without limit. In
the lensed 
case, the situation is more complicated because in the multiply imaged
regions the \clean{}ing is constrained by lens model. If these regions are of
comparable size to the resolution of the observations (or even smaller), the
residuals can still 
be reduced significantly below the noise level, especially in combination with
self-calibration (see below).

\begin{figure*}
\includegraphics[width=\textwidth]{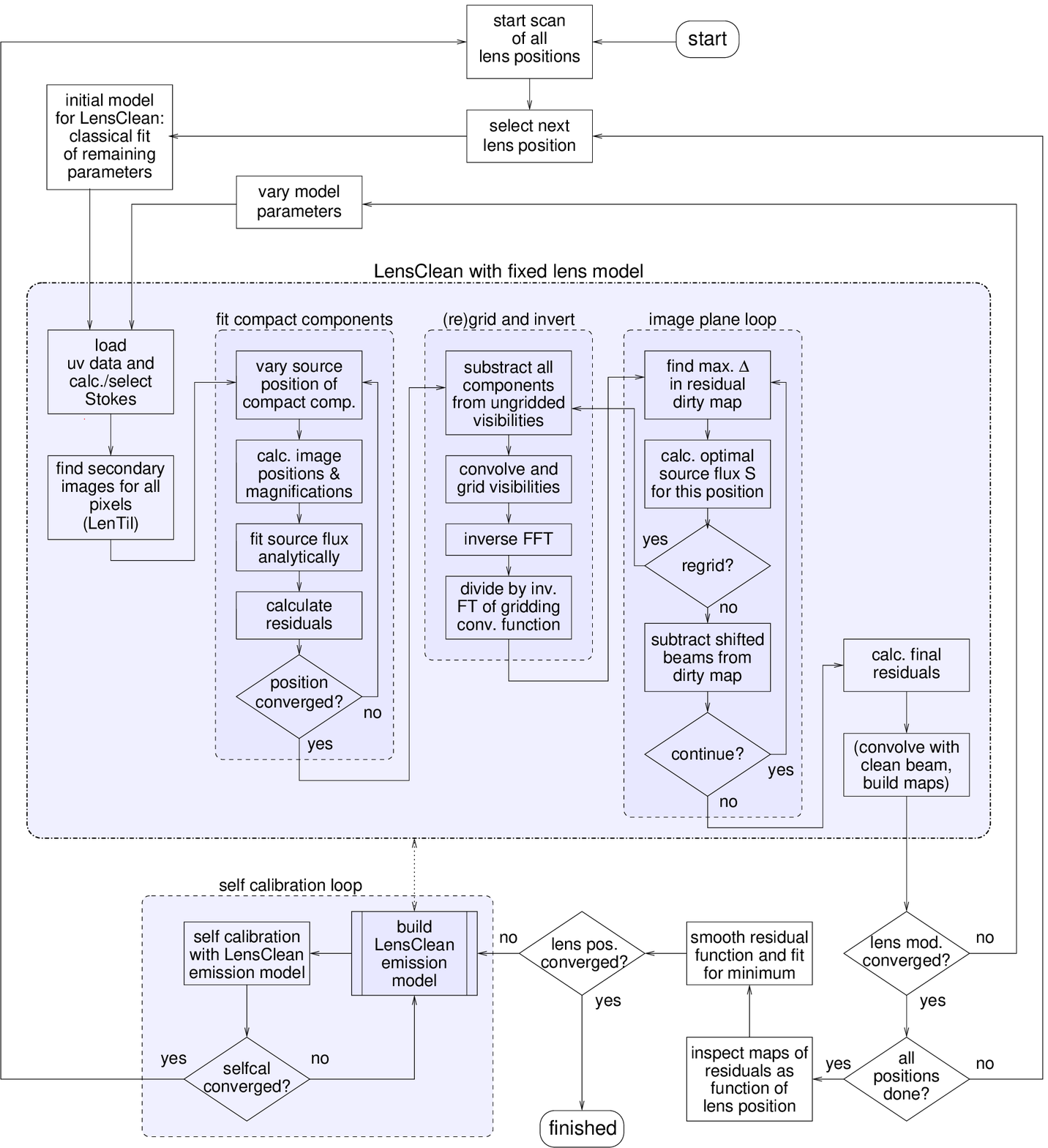}
\caption[Flow-chart of \lc]{Schematic flow-chart of the
  complete \lc\ algorithm as we used it for fitting 
  lens models for \BOii. Most important is the inner core (large shaded box)
  which builds an 
  emission model and calculates the residuals for a given fixed lens model. If
  required, maps of the lens and source plane are also produced. Built around
  this core is an algorithm to determine the optimal lens model, including a
  loop for the self-calibration of the data. Inside of the self-calibration
  loop the same \lc\ core is used to build an emission model.}
\label{fig:flow}
\end{figure*}

\subsection{Details of the inner core}

Figure~\ref{fig:flow} illustrates the structure of our implementation of the
inner \lc\ core in the large shaded box. In detail it works like this.
Before any \clean{}ing is done, we calculate for each pixel (`primary image')
in our image plane map area the corresponding source position and the
positions of all other `secondary images' which are produced for the same
source 
position by the lens. Finding all images for a
given source position for a SIEP lens model implies solving a quartic equation.
The one already known primary image can be divided out,
leaving a cubic equation to solve. This can and should be done
analytically with special care to avoid any errors in this step. Incorrect
image positions for even one pixel in the map will make the residuals
after \clean{}ing unusable and hence make model fitting
impossible. Errors for individual pixels will often allow more
freedom for the \clean{}ing process and thus reduce the residuals. The outer
loop of lens model fitting will happily stick at such models, leading to
wrong results.

When all images are found, the magnifications are calculated.
Sometimes it happens that a few pixels have very high
magnifications of the order of a few hundred. These can change the \lc\
residuals considerably which increases the numerical noise of the residual
function.  With artificial data we
established that the fits improved when such pixels (including their secondary
images) were excluded from \lc{}ing without introducing systematic
errors.

For each iteration, the optimal source position (on the regular grid required
by the FFT)
and flux density is found with the unbiased \lc\ method described
above. Scaled dirty beams are then subtracted in image space for each image
position. For the secondary images, which are not located at exact grid
positions, we distribute the flux over the four nearest pixels in a way which
is equivalent to bilinear interpolation. This is essential to avoid
large discontinuities in the residual function which would fool the outer
residual minimization with local minima.
Errors of the image space
representation are minimized by using a truncated Gaussian multiplied
by a sinc-function as gridding convolution function and by using small
pixels. Aliasing 
is only noticeable in the first few \clean\ iterations when the subtracted
flux 
density is still high.
After a number of iterations of this kind, all accumulated \lc\
components are subtracted at their exact positions from the ungridded
visibilities which are then regridded to get a new residual map
with gridding errors of all iterations up to then removed. This is known as
Cotton-Schwab algorithm in unlensed
\clean\  \citep{schwab84,cornwell99} but is much more important in \lc\
because of the secondary images which are not located exactly on the grid.
Working directly on the $uv$ space visibilities would be desirable but is
computationally prohibitively expensive because it would imply inversion of
the FT in each iteration for many image (or source) positions in order not to
miss the optimal position.

Because of the dominance of the bright components, it is essential to
keep any errors caused by these as small as possible. These residual errors
would otherwise hide any effects of the ring 
which are essential to obtain constraints for the galaxy position.
To avoid errors caused by the discrete nature of the pixels for the bright
images, we remove them partially before the normal iteration starts. This is
done by one \emph{ungridded} \lc\ iteration (model fitting with a free
continuous 
source position and flux and subtraction of the best fit afterwards)
at the beginning with a loop gain 
of nearly unity so that the compact image is removed without
affecting the smooth background of the ring. The ring itself will influence
this first ungridded 
step to some degree, which might produce bias effects in later stages. Tests
with 
artificial data showed, however, that these biases are too small to affect the
results seriously. Small errors in the positions of the bright
components will be corrected for in the later iterations anyway.

We did not enforce zero flux outside of user-selected windows because tests
with simulated data showed that this can introduce serious systematic shifts
of the residual minimum.

\subsection{Non-negativity constraints}

To reduce the freedom of the models without excluding physically realistic
models, a non-negativity constraint for the flux components would be desirable.
In the lensed case, combinations of positive and negative components in
regions of rapidly varying magnifications close to the caustics can produce
image plane brightness distributions which could not be reproduced by only
positive components with the same lens model. They therefore reduce the
residuals 
for incorrect lens models without changing them much for the best lens
model. With a non-negativity enforcing algorithm the accuracy of the lens
model could therefore improve considerably.

Well known approaches in \clean, like allowing only positive components,
stopping at the first negative component or deleting negative components
afterwards, are not able to find the best non-negative solutions because
negative components occuring during later stages of \clean\ are often only
needed to compensate for overestimated positive components in earlier
iterations. This happens especially if the weighting of the data is changed
between \clean\ iterations.
Disregarding these compensating components prevents \clean\ from finding an
optimal solution.
 Other methods like the \nnls\ algorithm\footnote{The acronym
  \nnls\ stands for 
  `non-negative least squares'. It finds the solution of lowest residuals under
  the non-negativity constraint. Its application to radio interferometry in
  the unlensed case was discussed by \citet{briggs_phd,briggs95}.} are able to
derive a true optimal non-negative solution in the unlensed case. 
In principle, \nnls\ can be extended for a lensed scenario in a
straight-forward way. Unfortunately, the numerical difficulties are much more
serious than in the unlensed case.

We recently found an alternative solution by modifying \clean\ to include a
non-negativity constraint in a strict way. It works by allowing negative
components only at positions where positive ones have been put before and only
with absolute fluxes smaller than the total combined flux at this position up
to that iteration. In this way negative components are allowed to modify
earlier positive ones, but negative total fluxes can never occur. This new
method is not implemented in standard software packages like \aips\ or
\difmap\ but is now part of our own code. Tests have shown that it can improve
the solution in the unlensed case significantly, especially if the emission
model is to be used for self-calibration. In the latter case, the well-known
negative features resulting from incorrect calibration can not be included in
the emission model so that they are removed later with self-calibration, at
least 
if there are large regions without emission in the map area so that the
non-negativity constraint can become active at all.

In the lensed scenario, the flux density from Eq.~\eqref{eq:S KNE} or
\eqref{eq:S unbiased} is used directly only
if it is either positive itself or if at least the sum with the already
existing components at the same position is not negative. Otherwise the
absolute value of a negative $S$ is decreased until the sum of components at
this position exactly vanishes. This procedure is followed for all pixel
positions and the resulting potential decrease of residuals $\Delta$ is
calculated from Eq.~\eqref{eq:R^2 2} with the modified value for the source
flux $S$. In the same way as without the non-negativity
constraint, the pixel with maximal $\Delta$ is then selected in this
iteration.

Experiments with this variant of \lc\
showed that it does indeed reduce the unwanted freedom of models but does on
the other hand introduce discontinuities and bias effects in the residual
function. These are much weaker than with the more standard methods of
rejecting negative components, but they are still significant. Because these
effects have to be avoided in any case, we generally did not exclude
negative components. The only exception is the brightness model used for
self-calibration (see below). But even here the inclusion of negative
components would not have changed the results noticeably.

The reason for the current failure of non-negativity enforcing versions of
\lc\ probably 
lies in the fact that total negative components are often necessary to
compensate for small errors originating from other effects, e.g.\ the gridding
of the model. These 
errors can change with variations of the lens model so that the increase of
residuals caused by not allowing negative components produces lens model
dependent bias effects. A better understanding of these effects is highly
demanded and it is our hope that a better algorithm including the
non-negativity constraint without introducing other problems can be developed
in the future.

\subsection{The outer loop: Fitting the lens model}

After \lc\ found the best source brightness model for a given set of lens
model parameters, the residuals from the inner loop are used to find the best
lens model in the outer loop (see Fig.~\ref{fig:flow} for an illustration).

Even with all the precautions discussed before, the residuals are no
absolutely smooth function of 
the lens model parameters, making it difficult to fit the lens
model. We use the downhill simplex method \citep[cf.][]{press_etal} because
more sophisticated minimization methods which rely on smooth quadratic minima
did not prove to be 
superior in this case. Simultaneous \lc\ fits of all five free parameters
are dangerous in our case; they often get stuck at local minima.
This is a result of the very different magnitude of the eigenvalues of the
matrix describing the residual function near its minimum. Three of the
lens model parameters 
could be fitted with the very bright compact images alone, leading to three
very 
high eigenvalues. For the remaining two, the position and structure of the much
weaker and less compact structures in the ring have to be used, causing two
much smaller eigenvalues. This combination of very high sensitivity in certain
directions with very low sensitivity in other directions is a highly
problematic case for any minimization method. Finding the global minimum of
a long, narrow and bent valley is never easy, especially if numerical noise
(not to be confused with the noise of the observations) causes additional
fluctuations.

To be absolutely 
sure to find a global minimum in the allowed parameter range, we
scan all realistic values of the lens position $\vc z_0$ and fit only the
remaining three model parameters (lens mass scale $\alpha_0$ and ellipticity
$\epsilon_x$ and $\epsilon_y$) for each fixed $\vc z_0$.
A number of three parameters could be fitted with the information from the two
strong components alone with very high accuracy. We therefore avoid using the
low eigenvalues provided by the ring for the minimization of residuals and
achieve a highly improved stability.
The ring still shows its effects in the
best residuals as a function of $\vc z_0$, of course.
To assist the \lc\ model fitting, we start with a classical lens model fit
for each $\vc z_0$ which is already very close to the final model.
The combined scanning/fitting approach is nevertheless numerically very
expensive and at the limit of what can be done with a cluster of
modern PCs or workstations. It does on the other hand allow the estimation of
confidence regions from the maps of residuals as function of $\vc z_0$ and
provides an invaluable diagnostic to detect possible 
numerical problems. Only if the algorithm works optimally, do the residuals
show a smooth quadratic minimum. To reduce the remaining
numerical noise, smooth polynomial functions are fitted to the residual
function $R^2(\vc z_0)$ to determine the minimum and confidence regions.
A similar fit for the remaining lens model parameters evaluated at the
residual minimum provides the final optimal lens model.

We also tried to separate the effect of the two compact components from the
effect of the ring. One attempt was to fit the other parameters for given $\vc
z_0$ classically and use \lc\ only to calculate the final residuals for this
model. This method, if successful, would accelerate the complete model fitting
by a factor of $\sim 100$.
Unfortunately the residuals from the two compact components are so strong,
although we tried to remove their influence by several methods, that they lead
to serious 
systematic errors in the final results. These approaches were therefore not
used to produce the results presented in \papii.

In order to correct for possible changes of the flux ratio A$/$B caused by
variability in combination with the time-delay or by propagation effects, we
also fit lens models to 
modified data sets where we either added or subtracted flux at the position of
one of the components to compensate for the effect. Another approach is to
artificially change the amplification\footnote{We change the
  \emph{amplification} for unresolved components but not the
  \emph{magnification} of the lens itself.} of the lens model by some amount
in a 
small region around one of the bright images. We found that both methods
lead to very similar results. The effect on the best lens models in the case of
\BOii\ will be discuss in \papii.

\subsection{Self-calibration}

We started our \lc\ algorithm with a calibrated data set which was first used
to make a map with unlensed \clean\ and used the self-calibration from this
procedure. 
After finding a best fitting lens and source model for the given data, we
used this model (now with positivity of components enforced) to self-calibrate
the data set.  It is important that the same weighting is used
for the self-calibration as for the \lc\ model fitting. For uniform weighting,
the data set was re-weighted to be able to use \difmap\ for the
self-calibration. A moderate number of alternating \lc\ runs and 
self-calibration 
steps where used for the (fixed) lens model.
We then started with these
recalibrated data from the beginning (see Fig.~\ref{fig:flow}). It showed that
only a few 
iterations of \lc\ lens model fitting and self-calibration were necessary to
get a stable result.

In the self-calibration we allowed independent  phase and magnitude
changes for each 1~min integration. This may introduce too much
freedom but has the advantage that (if the same procedure is applied to
artificial data sets) the real and simulated data are guaranteed to have the
same level of calibration so that results of the two can be compared with
confidence. 

In order to test the robustness of the procedure we also started
with very badly calibrated data and with
self-calibrating with an incorrect lens model. After a few iterations the
fits always converged to the same best solution which established the
stability of this method.

It is not possible to use self-calibration as part of the innermost \lc\ loop
as it is often done with \clean, because the comparison of residuals of
different self-calibrations introduces an extreme level of numerical noise
which completely hides the effects of different lens models. The method would
also be numerically too expensive to be applied on a regular basis.
The extensive numerical tests with real and simulated data proved that using
self-calibration only after an optimal lens model is found is sufficient to
remove any initial calibration errors and to
find the best solution.

\subsection{Goodness of fit, error statistics}

Several stopping criteria for the \lc\ iteration have been
discussed by \citet{kochanek92}. We decided to use the simple scheme of taking
a 
fixed number of iterations (about 500 to 5000), using the remaining
residuals $R^2$ as a 
direct measure for the goodness of the fit. In the outer loop, the lens
model parameters are then varied to find a minimum of $R^2$. We
decided to use the $uv$ space residuals $R^2$ instead of other measures for
the residuals for 
the outer loop in order to get a consistent solution for the lens model and
source structure.
When computing residuals in $uv$ space, we also avoid the
problem of correlated errors in image space.

Different
weighting schemes were tested with artificial data sets with the
result that the high sensitivity for small scale structures provided
by uniform weighting helps in getting useful constraints for the models in the
initial stages of self-calibration. 
Interpretation of the residuals is simpler for natural weighting
of course, because in this case $R^2$ is equal to $\chi^2$ and the complete
algorithm is equivalent to a maximum likelihood fit of the lens mass model and
source brightness distribution.

For real statistical weights of $1/\sigma_j^2$ and applied weights
$w_j$, the expected value of $R^2$ for the correct lens and emission model is
\begin{equation}
<R^2> = \sum_j w_j \sigma_j^2
\label{eq:R2}
\end{equation}
and the standard deviation
\begin{equation}
\sigma_{R^2} = \sqrt{2\sum_j w_j^2 \sigma_j^4} \rtext{.}
\label{eq:sigma2}
\end{equation}
Usually the effective number of parameters of the \emph{emission model}
(estimated by 
the number of non-overlapping beams in the map area) is much smaller than the
number of visibilities so that the implicit fit of the emission model does not
change 
the error statistics significantly. The residuals will be reduced by some
amount, but this 
reduction is (almost) the same for all lens models and does not influence the
fit of the lens model. The only concern could be that the effective number of
emission model parameters changes with the lens model via changes of the area
of 
doubly and quadruply imaged regions, in which the emission model has less
freedom than in singly imaged regions. The effect of this has been estimated
with artificial data consisting of only noise. Differences of the residuals
for different lens models are a direct measurement of this possible bias.
From these simulations we
learned that the shift of the best lens model originating from this effect is
much smaller than the statistical errors and can thus be neglected in the
following. 

We now assume that the emission model has been fitted for each lens model and
discuss the residuals without worrying about the emission models.
The best fitting \emph{lens model} will have residuals which are smaller than
the expectation from Eq.~\eqref{eq:R2} by $\Delta R^2$ 
(it has by definition the smallest residuals of all models).
If the number of fitted parameters is much smaller than the number of
visibilities, $\Delta R^2\ll R^2$ and the residuals can be used directly to
judge the goodness of fit.
For $\chi^2$ statistics (natural weighting: $w_j=1/\sigma_j^2$), mean and
standard deviation of the residual minimum are $\nu$ and 
$\sqrt{2\nu}$, respectively, where the number of degrees of freedom
$\nu$ is the difference of the number of visibilities (real and
imaginary part counted separately) and the number of fitted
parameters. The difference $\Delta \chi^2$ also follows a $\chi^2$
distribution with $\nu$ given by the numbers of fitted parameters and can be
used to determine confidence limits.
For non-natural weighting schemes, this simple
interpretation is not possible. In this case $\nu$ does not only
depend on the weights and the numbers of visibilities and parameters
but also on the model and the data themselves.
We then can still use the best $R^2$ to judge if our fit is acceptable
but we cannot use differences $\Delta R^2$ to the best $R^2$ to determine
confidence limits in the rigorous way in which it is possible for natural
weighting from the $\chi^2$ distribution.

In \citet{phd} we presented an analytical approximation which makes it
possible to \emph{estimate} confidence regions also for general weighting. The
result 
applied to our case says that the difference of the residuals for the best
fitting and true model is expected to be
\begin{equation}
\left<\Delta R^2\right> \approx \frac{\sum_j w_j^2 \sigma_j^2}{W}
\label{eq:DR2}
\end{equation}
for one (lens-)model parameter. By analogy to natural weighting where this
value becomes unity, we scale this residual difference with the normal limits
from the $\chi^2$ distribution to obtain different confidence limits for an
arbitrary number of parameters.
\iffalse
 Monte Carlo simulations showed that this
approach indeed provides a good approximation to the true statistical
uncertainties of lens model parameters fitted with \lc.
\fi

\begin{figure}
\includegraphics[scale=0.8]{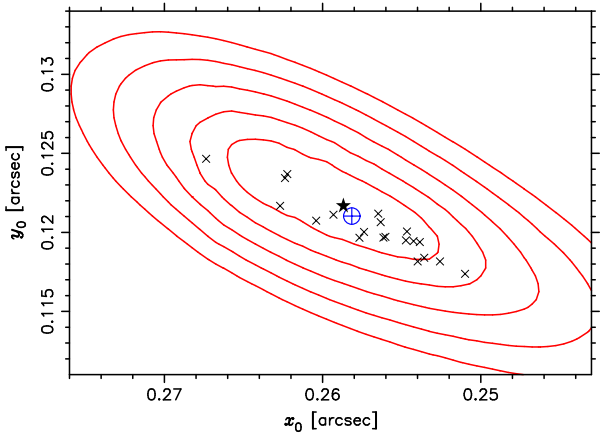}
\caption[Monte Carlo results]{Results of Monte Carlo
  simulations for the 15\,GHz VLA data (Stokes I) showing residuals as a
 function of lens position. We picked
 one of the runs with a minimum (star) close to the model used to
 build the data (crosshair) for the
contour  plot (confidence limits of 1, 2, 3, 4, 5$\cdot \sigma$). 
 The other Monte Carlo results are shown as crosses.

}
\label{fig:mc}
\end{figure}

To detect possible systematical errors and to check the accuracy of the
confidence limit estimates for uniform weighting, we performed a small
number (21) of Monte Carlo simulations. Even though  we did not include
self-calibration to save computing time, a total of about two years of CPU
time on modern PCs was needed for these simulations\footnote{We parallelized
  most of the calculations using normal workstation type PCs. With a typical
  number of 25 computers we need about one month of real time for a simulation
  like this.}.
We took a model for the lens and brightness of the source from earlier stages
of our \lc\ fits of the \BOii\ data to produce an artificial data set and used
the same algorithm (without self-calibration) as for the real data with
uniform weighting. The source 
model does not reproduce the real data exactly so the results are not meant to
be compared with the results of the real fits. The idea is instead
to compare the distribution of residual minima with the expectations from the
residual differences and error statistics.
The results in Fig.~\ref{fig:mc} show that the agreement is indeed quite
good, although the simulated distribution seems to be somewhat flatter than
the approximated estimate. Of the 21 runs a number of 17 (81 per cent) is
within the expected 
$1\,\sigma$ region and all are within $2\,\sigma$. Since error statistics
expects only 68 per cent inside the $1\,\sigma$ region we conclude that our
error estimates are at least not excessively overoptimistic.

Despite the cautionary note above, the size of the
simulated confidence regions is also very similar to the real data results
which are presented in \papii.

\section{Non-isothermal models: LenTil}

We explained before that \lc\ relies on a very robust method to solve the lens
equation i.e.\ to find all images for a given source position. Because of this
we have done most of our work with isothermal elliptical potentials for which
this can be done analytically.
To be able to use more general lens models with \lc, we developed a new
algorithm which can invert the lens equation for any lens model for which the
deflection angle can be calculated  as a continuous  function of the image
position (with the possible exception of known singularities).

\begin{figure*}
\includegraphics[width=0.9\textwidth]{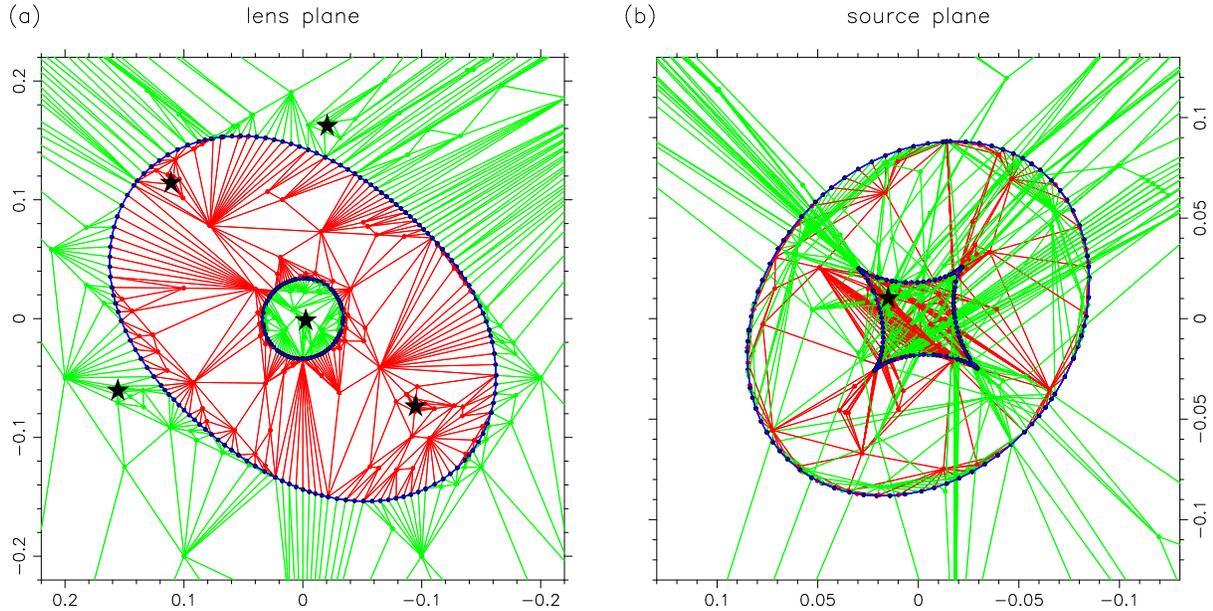}
\caption[Image search with \lentil]{Image search with the \lentil\
  algorithm. In addition to the initial tiling, this plot includes the
  subtilings performed in the search of images for one source
  position. (a) Lens plane with critical curves and five images (stars),
  (b) source plane with caustics and source position (star).}
\label{fig:lentil}
\end{figure*}

Our algorithm (called \lentil\ for `lens tiling') is based on ideas similar to
those used by \citet{keeton01a,keeton01b} in his software package. \lentil\
is adapted especially for the use with \lc\ which means it has to be
extremely reliable (failure in less than one of $10^8$ cases) while still
being sufficiently fast to be able to invert the lens equations for all pixels
without increasing the already high computational demands of \lc\ too much.
\lentil\ is used for very many source positions for each lens model so some
overhead can be accepted to prepare the calculations for each model. Including
this overhead, the total time for one inversion is of the order a few
milli-seconds. 

The basic idea is the following. The lens plane is subdivided in a large
number of triangular tiles which are then mapped to the source plane with the
given lens model. There it can be tested in which of the source plane tiles
the source position is located. Other tiling methods would then use the lens
plane position of these tiles to start standard numerical root finding
algorithms from there. This approach is well suited for standard lens
modelling where the failure in a few cases does little harm but this is
not 
acceptable for \lc. Our algorithm therefore uses the tiles in which the source
is located to start a subsequent subdivision until the required accuracy is
reached. During this subdivision the source will often leave the initial tile
and be located in a neighbouring tile after a subdivision step. In these cases
the subdivision process continues with these new tiles.
Special care is needed to assure convergence of the subdivision process in
all cases. Sometimes the subtiles become degenerate which has to be avoided
because it prevents convergence. The tests of whether or not the source is
located 
within a tile have to be done in a special way to lead always to consistent
results even if the source is located exactly (in a numerical sense) on one of
the bordering edges of a tile. This is difficult when the source lies not only
on one of the edges but exactly on one of the vertices (i.e.\ on at least two
edges). These possible problems have to be detected and must in bad cases be
corrected by shifting the source by a very small amount slightly above the
numerical resolution but far below the required accuracy, and restarting the
algorithm. 

In the preparation phase, which has to be performed once for each lens model,
the initial tiling is built. In this phase it is extremely important to detect
all critical curves which separate regions of different parities because these
regions have to be treated separately in the later subdivision
stages. Critical curves have to be sampled sufficiently densely to
avoid loosing very close multiple images with a high magnification. Some
special care is also needed to treat different kinds of singularities of lens
models (pointmass-like ones with diverging and SIS-like
ones with finite deflection angles) correctly.

We start by covering the area of interest with one very large
triangular tile. We then divide this initial tile by adding new vertices very
close to all the (known) singularities of the lens model. In the models used
with \lc\ so far, we only had one singularity per model. From the kind of
singularity we infer the number of critical curves that should enclose
it. After searching for the critical points on a straight line drawn from the
singularity outwards, we put new vertices in all the domains of different
parities. Then we subdivide the tiles until their sides are all below a
preselected limit and until the critical lines and `cuts' are sampled
sufficiently densely.

In all these and the following subdivisions we always take care that
no side of a tile ever crosses a critical curve, i.e.\ that no side connects
regions of different parity, by introducing additional vertices at such
crossing. This is important to assure that the tiles do (in
the limit of infinite subdivision) project to the source plane with a well
defined parity. Otherwise it could happen that some regions of the source
plane are missed or others are sampled several times. This would result in
missing images or additional phantom images. Both have to be avoided.

Figure~\ref{fig:lentil} shows a typical tiling in the lens and source plane
with five images found by the subsequent subdivision.
After an extensive testing phase, the \lentil\ algorithm is now in a state in
which it can be used for \lc. Depending on the lens models, it can still
double the CPU time needed compared to \lc\ with the analytical SIEP inversion
but this is regarded as acceptable.

The \lentil\ code is refined continuously and
it is therefore premature to make it publically
available.
More details of the implementation can be
found in \citet{phd}.

\section{Source reconstruction}

Although the pointlike \clean\ components are an optimal representation of the
data in the sense of a maximum likelihood model, they usually represent the
true source maps very badly because they contain signals with very high
spatial frequencies which are not measured directly because of the limited
$uv$ range of the observations. These high frequency parts are highly
uncertain and should therefore be reduced to produce the final maps. The
standard regularisation method with \clean\ is to convolve the collection of
components with a Gaussian `\clean\ beam' whose size is chosen in a way to
resemble the dirty beam in its central parts. This is a sensible approach
although it has its shortcomings and there is no rigorous mathematical
foundation for it. In practice the standard \clean\ beam convolution is used
on a regular basis and leads to satisfying results. We therefore want use
the same idea generalized to the lensed situation to produce source plane maps
with \lc.

The approach presented by
\citet{kochanek92} uses circular beams and chooses the size so that
they just cover the intersection of the projected single image
beams of all corresponding images. With this approach the combined beam is in
each direction 
larger or equal to the smallest of the single beams and does not
introduce spurious small scale features.
This procedure has its justification for equal magnifications,
but becomes very inaccurate for high magnification ratios. In this
case the projected beams should be weighted in some way according to the
lens plane flux densities which are proportional to the magnifications of the
images. Images with very low magnification should contribute less to the final
beam than images with a high magnification.

Before we can translate the concept of a beam to the source plane, we have to
understand its meaning in the lens plane in the context of the linear least
squares problem which is solved by standard \clean. We formulate the problem
in an algebraic way which can then be extended easily.
The general problem of this kind can be written like this:
\begin{equation}
\vc y = \mat A \, \vc x + \text{noise}
\label{eq:y=Ax}
\end{equation}
Here $\vc x$ denotes the vector of model parameters (brightness distribution
in our case),
$\vc y$ is a vector describing observations (visibilities in our case)
and the linear function connecting the two is written as a matrix $\mat
A$ (the Fourier transform here). This equation is the
general version of Eq.~\eqref{eq:ft radio} but the brightness distribution $I(\vc
z)$ 
is written as a discrete vector $\vc x$ and the visibilities $\ft I_j$ as
vector $\vc y$. With a weighting matrix $\mat W$,
the residuals for observations $\vc y$ and model $\vc x$ are
\begin{equation}
R^2 = \transp{(\vc y-\mat A\vc x)} \, \mat W \, (\vc y-\mat A\vc x) \rtext{.}
\nt
\end{equation}
In our case the matrix $\mat W$ consists only of the visibility weights as
diagonal elements.
The residuals are minimal if the derivative with respect to $\vc x$ vanishes,
which leads to the equation
\begin{equation}
\mat B \, \vc x = \vcID \rtext{,}
\label{eq:conv}
\end{equation}
where $\mat B$ denotes the matrix of the dirty beam $(\mat B)_{jk}=B
(x_j-x_k)$ and $\vcID$ the vector of the dirty map $(\vcID)_j=\ID(x_j)$:
\begin{align}
\mat B &= \frac{1}{W} \transp{\mat A}\mat W\mat A 
\label{eq:B AWA}
 \\
\vcID &= \frac{1}{W} \transp{\mat A}\mat W \, \vc y \nt
\end{align}
The normalization with $W=\trace \mat W$ is used by convention but is not
necessary. In this way the dirty map resembles physical units and the dirty
beam has a central value of unity.

If a lens is introduced in the problem, we can write the brightness
distribution in the lens plane as a linear function of the brightness
distribution of the source plane:
\begin{equation}
\vc x = \mat L \, \vc x\sub s
\nt
\end{equation}
The matrix $\mat L$ describes the lens effect. If the vectors $\vc x$ and $\vc
x\sub s$ are used to describe a collection of $\delta$ components, $L_{kk\sub
  s}$ is the magnification of the image $\vc z_k$ if it is mapped to the source
position ${\vc z\sub s}_{k\sub s}$ and 0 otherwise. Of course the
representation of 
$\vc x$ and $\vc x\sub s$ has to be complete, i.e.\ each component of $\vc x$
must correspond to one component of $\vc x\sub s$ and each component of $\vc
x\sub s$ must correspond to one or more images in $\vc x$.
Writing Eq.~\eqref{eq:y=Ax} with $\vc x\sub s$ instead of $\vc x$, we
see that the effect of the lens is to replace the matrix $\mat A$ by a
modified matrix $\mat A\sub s=\mat A\mat L$. The `source plane dirty beam' can
then be defined as
\begin{equation}
\mat B\sub s = \transp{\mat L} \mat B \mat L
\nt
\end{equation}
analogous to the lens plane dirty beam in Eq.~\eqref{eq:B AWA}.
This source plane dirty beam has,
however, not the same properties as $\mat B$. Whilst $\mat B$ is translation
invariant so that Eq.~\eqref{eq:conv} can be read as a convolution, this is not
true for $\mat B\sub s$ in the general case.
Explicitly, the source plane dirty beam can be calculated using the definition
of $\mat  L$:
\begin{equation}
B\sub s (\vc z\sub s,\vc z'\sub s) = \sum_{\substack{\vc z(\vc z\sub s)\\
    \vc z'(\vc z'\sub s)}} \mu (\vc z) \mu (\vc z') B (\vc z,\vc z')
\label{eq:dirty beam source 1}
\end{equation}
In this sum $\vc z$ and $\vc z'$ run over all images of the source
components $\vc z\sub s$ and $\vc z'\sub s$, respectively.
For our first attempts to produce a map of the source plane of \BOii, which
will be presented in \papii, we
introduce some approximations. We assume that the images are well separated,
i.e.\ that their respecting dirty beams do not overlap (only the diagonal
elements of the sum in Eq.~\eqref{eq:dirty beam source 1} contribute) and that the
magnification close to the individual images is not varying significantly over
an area corresponding to the beam.
With these approximations it is possible to express distances in the lens plane
$\Delta\vc z$ by the corresponding distance in the source plane $\Delta \vc
z\sub s$ using the local magnification matrix $\mat M$:
\begin{equation}
B\sub s (\Delta\vc z\sub s) = \sum_{k} \mu_k^2 \, B (\mat M_k \Delta \vc z\sub s)
\label{eq:dirty beam source 2}
\end{equation}
The index $k$ runs over all images for the given source position. For small
$\Delta\vc z\sub s$ this source plane beam is approximately translation
invariant but it depends on the source position on larger scales.

Now we use the standard approximation of a Gaussian for the lens plane dirty
beam and write
\begin{equation}
B(\Delta\vc z) = \expa{-\transp{\vc z} \mat G \vc z/2} \rtext{.}
\label{eq:B G}
\end{equation}
The matrix $\mat G$ determines the parameters of the Gaussian and can be
calculated from the major and minor axis FWHM\footnote{Full width at half of
  the maximum.} $a$ and $b$ and 
position angle $\phi$ of the beam. The same parametrization as discussed in a
different context in 
Appendix~\ref{II-sec:app shapes} of \papii\
can be used here. If the
parameters are used as defined before, the beam matrix is $\mat G=(8\ln2) \,
\mat E^{-1}$. 
With the properties of the Fourier transform it is easy to show that $\mat G$
is proportional to the second moment matrix of the $uv$ coordinates and can
thus be calculated easily.

Now we can fit an effective total Gaussian to the central part of the  source
plane beam from Eq.~\eqref{eq:dirty beam source 2}. This is done by expanding the
exponential in Eq.~\eqref{eq:B G} to first order and write the sum again as
exponential for which it 
is the first order Taylor expansion. The resulting beam is again a Gaussian,
now with central value $\sum \mu_k^2$ and with parameter matrix
\begin{equation}
\mat G\sub s = \frac{\sum\limits_k \mu_k^2 \, \mat M_k \mat G \mat M_k}{\sum\limits_k \mu_k^2} 
\rtext{.}
\label{eq:beam source 1}
\end{equation}
This is a weighted mean of the individual backprojected matrices $\mat M_k
\mat G \mat M_k$. For very different individual source plane beams (i.e.\ for
very different magnifications) the combined beam is dominated by the smallest
individual beam which has the highest magnification. This means that the
effective source plane resolution can be improved beyond the lens plane
resolution by the lens.

Beams in standard \clean\ are normalized to have a central value of
unity. This leads to maps in physical units per beam (e.g.\
$\mathrm{Jy}/\mathrm{beam}$). For resolved sources this can be transformed
to units of surface brightness (e.g.\ $\mathrm{Jy}/\mathrm{arcsec}^2$)
easily. In our case, the beams are not constant 
but depend on the position in the source plane. If the same normalization
would be used in this case, the resulting map would have varying units over
the field and would not conserve total flux. Even for well resolved sources
with constant surface brightness, the
values of the map would depend on the position which is clearly not desirable.
We therefore normalize the beam to have not a constant central value but a
constant total flux of unity. This leads to maps in units of surface
brightness and preserves total flux.
The final source plane beam is then
\begin{equation}
B\sub s(\Delta\vc z\sub s) = \frac{\sqrt{\lvert\mat
    G\sub s\rvert}}{2\pi}\expa{-\transp{\vc z\sub s} \mat G\sub s \vc z\sub
    s/2} \rtext{.}  
\label{eq:beam source 2}
\end{equation}
The source plane reconstruction with a beam following Eq.~\eqref{eq:beam
  source 1} and \eqref{eq:beam source 2}
is certainly not optimal. The approximations used do not work very close to
caustics and the whole procedure is somewhat arbitrary. On the other hand it
is 
based directly on the standard \clean\ beam convolution which is known to work
well. Any more
optimal method will probably be an integral part (in the sense of
regularisation) of the algorithm and not be 
applied at the end once the best brightness model is known. Investigations of
better approaches could therefore start with the simpler unlensed problem and
be generalized from that.

It is now possible to use the reconstructed source plane beams and map them
back to the image plane to produce a superresolved image plane map of the lens
system. In singly imaged regions, this superresolved map is equal to a normal
\clean\ map because projecting the beam forth and back does not change it. It
is only the combination of several beams in multiply imaged regions which
provides the opportunity to improve the resolution in the lens plane map.

In \papii\ we show reconstructed maps of the unlensed source of \BOii\ on
different scales as well as standard and superresolved lens plane maps.

\section{Discussion}

Multiple images of lensed extended sources do potentially provide far better
constraints for lens mass models than simple multiply imaged point sources
because they probe the lensing potential not only at a small number of
discrete positions but over wide areas of the lens plane.
In order to utilize these constraints, it is necessary to fit not only for the
mass model of the lens but also (implicitly or explicitly) for the brightness
distribution of the source. This makes the model fitting for extended sources
much more difficult than for compact sources which can be described by a
position and flux density alone.
We showed how the task can be performed using an improved version of the \lc\
algorithm. In an inner loop, the brightness model of the source is optimized
for a given lens model, while an outer loop uses the residuals from the inner
loop in order to determine the optimal lens model parameters.

As a test case we used the lens \BOii\ where the beautiful Einstein ring shown
by radio observations potentially provides constraints to determine the
position of the lens with sufficient accuracy.
Surprisingly, variants of \lc\ have been used only for very few cases
before.
This is partly a result of the high numerical
demands. In the case of \BOii\ we furthermore showed that the algorithm in its
original form is not able to produce reliable results because of the
dominance of the bright compact images which hide the more subtle effects of
the ring.
Special measures are required to produce accurate results and this led
(amongst other improvements) to the implementation of an important new 
concept in \lc\ which is the unbiased selection of components.

We performed extensive tests both with real and simulated data to gain
confidence in the accuracy and reliability of our \lc\ variant. Finally the
algorithm was applied to a VLA data set of \BOii\ resulting in good
constraints for all parameters of an isothermal lens model including the lens
position. These results are presented in the accompanying \papii.

To be able to use non-isothermal models with \lc, we had to develop a new
technique to solve the lens equation for very general models fast and (above
all) very reliably. The new \lentil\ method has been tested and allowed first
applications of \lc\ for non-isothermal lens models.

A secondary result of \lc\ is a brightness distribution model for the
source. We presented a new approach using the concept of a `source plane dirty
beam' to produce a source plane map from the
\lc\ components utilizing the lens magnification to resolve small substructure
in the source which could not be seen otherwise.

Further improvements of \lc\ are possible. Most important 
seems to be the inclusion of a non-negativity constraint for the brightness
distribution in a way that avoids the problems of available approaches.
Significant
improvements in the discrimination between `good' and `bad' lens models are
expected from such a method.
We recently found a very simple way to include non-negativity constraints in
\lcc\ by allowing negative components only at positions where positive
components have been accumulated before and only up to a maximal negative flux
which cancels the positive components. This approach seems to work very well
in unlensed \clean\ but is still not satisfactory in \lc.

Although it is not important for the lensing
aspect, we are also working on new methods to reconstruct the unlensed source
plane. These methods would optimally be implemented as regularisation in the
algorithm directly instead of smoothing afterwards.

In the future the application of \lc\ for several lens systems with extended
radio structure will allow a systematic investigation of lens mass
distributions. This will not only help in cosmological applications but also
provide invaluable information about the lens galaxies themselves. No other
method would be able to 
study the mass distributions of high redshift galaxies accurately and
independent of dynamical model assumptions.

\section*{Acknowledgments}

It is a pleasure to thank Andy Biggs and Ian Browne for continuous support and
encouragement. Without their help in learning the theory and practice of radio
interferometry, this work would not have been possible. The anonymous referee
is thanked for a very helpful report.

The author was funded by the `Deutsche Forschungsgemeinschaft', reference
no.\ Re~439/26--1 and 439/26--4; European Commission, Marie Curie Training
Site programme, under contract no.\ HPMT-CT-2000-00069 and TMR Programme,
Research Network Contract ERBFMRXCT96-0034 `CERES'; and by the BMBF/DLR
Verbundforschung under grant 50~OR~0208.

\newcommand{\mnras}{\mbox{MNRAS}}
\newcommand{\baas}{\mbox{BAAS}}
\newcommand{\apj}{\mbox{ApJ}}
\newcommand{\aj}{\mbox{AJ}}
\newcommand{\aaps}{\mbox{A\&AS}}
\newcommand{\aap}{\mbox{A\&A}}

\bsp

\end{document}